\documentclass[mathleft,fleqn,%
]{an}
\usepackage{graphicx}
\usepackage[varg]{txfonts}
\usepackage{natbib}
\bibpunct{(}{)}{;}{a}{}{,}
\begin{document}

\Pagespan{1}{}
\Yearpublication{2014}%
\Yearsubmission{2014}%
\Month{0}%
\Volume{999}%
\Issue{0}%
\DOI{asna.201400000}%

\title{Impact of rotation on stellar models}

\author{Georges Meynet\inst{1}\fnmsep\thanks{Corresponding author:
        {georges.meynet@unige.ch}}, Andr\'e Maeder\inst{1}, Patrick Eggenberger\inst{1}, Sylvia Ekstrom\inst{1}, Cyril Georgy\inst{1}, Cristina Chiappini\inst{2}, Giovanni Privitera\inst{1,3}, Arthur Choplin\inst{1}
}
\titlerunning{Impact of rotation}
\authorrunning{Meynet et al.}
\institute{
Department of Astronomy of the Geneva University, CH-1290, Versoix, Switzerland
\and 
Leibniz-Institut fuer Astrophysik Potsdam, An der Sternwarte 16, D-14482 Potsdam, Germany
\and 
Istituto Ricerche Solari Locarno, via Patocchi, 6605, Locarno-Monti, Switzerland}

\received{XXXX}
\accepted{XXXX}
\publonline{XXXX}


\abstract{
After a brief recall of the main impacts of stellar rotation on the structure and the evolution of stars, four topics are addressed: 1) the links between magnetic fields and rotation; 2) 
the impact of rotation on the age determination of clusters; 3) the exchanges of angular momentum between the orbit of a planet and the star due to tides; 4) the impact
of rotation on the early chemical evolution of the Milky Way and the origin of the Carbon-Enhanced-Metal-Poor stars.}
\maketitle

\section{Impacts of rotation on stars}

Rotation has many impacts on the structure and the evolution of stars \citep[see for instance the recent review by][]{MM2012}, either single or in close binary systems.
Among the most significant physical effects due to rotation, we can mention 
\citep[see the book by][for a detailed discussion]{maederlivre09}:
\begin{itemize}
\item Rotation modifies the stellar structure equations by adding a centrifugal acceleration term in
the equation describing the mechanical equilibrium \citep[see e.g.][]{Meynet1997}.
\item It deforms the outer layers of a star that will no long shines with a uniform brightness \citep{Maeder1999, Rieutord2007}. 
\item It produces stellar wind anisotropies \citep{MD2001, Georgy2011}.
\item At the critical velocity, it induces mechanical mass losses at the equator \citep[see e.g.][]{Ekstrom2012}.
\item Rotation activates meridional currents and many instabilities like shear turbulence
that transport chemical species and angular momentum in the stellar interior \citep{Zahn1992}.
\end{itemize}
These effects have many consequences on stellar evolution:
\begin{itemize}
\item The evolutionary tracks and the lifetimes for both single stars and stars in close binaries are changed by rotation \citep{Brott2011, Ekstrom12}. The most extreme cases are obtained
when rotational mixing triggers a homogeneous evolution \citep{Maeder1987, Song2015, Szecsi2015}.
\item Rotation produces changes of the surface abundances \citep[see e.g.][]{Przy2010, MaePrzy2014}.
\item At the latest stages of the evolution of stars, rotation may have a significant effect on how stars explode in a core collapse
supernova event \citep[see the review by][and references therein]{Janka2012}.
\item The properties of the compact stellar remnants are also affected by rotation \citep{Heger2005, Suijs2008}.
\item Rotation has a deep impact on nucleosynthesis \citep[see the review by][and references therein]{Chiappini2013}.
\end{itemize}
Thus when rotating stellar models are used in population synthesis models, or when yields of rotating models are used
in models for the chemical evolution of the galaxies some important changes are expected with respect to the result of analogous models
using the outputs of non-rotating models. In the present paper, we focus on four topics: 1) the interplay between rotation and magnetic field;
2) How rotation does affect the isochrones; 3) How rotation does affect the process of planet engulfment; 4) and finally, how rotation does impact
stellar nucleosynthesis at very low metallicity.

\section{Rotation and magnetic fields}

Rotation of convective zones activates a dynamo. The solar magnetic field is generated in that way \citep{Parker1955}.
\footnote{See also the historical account in \citep{Stenflo2015}}. 
Some authors have also suggested that differential rotation in radiative layers may
activate a dynamo \citep{Spruit02}. 

Magnetic fields, at their turn can affect rotation. A surface magnetic field, whatever its origin, fossil or produced by a dynamo, produces a magnetic braking when
the material composing the stellar winds is funneled along the magnetic lines \citep{Kawaler1988, Town2010}. An internal magnetic field attached to the stellar cores, may also induce
a coupling between the core and the adjacent radiative layers, participating thus to the transfer of angular momentum from the core to the outer regions \citep{MMIV2014}.

Rotating models can be computed without any magnetic field, or with one or a few of the effects mentioned above. 
It is important to realize that when magnetic fields are involved, the outputs of the stellar models can be very different compared to analogous models
without any magnetic field. The large grids of rotating stellar models can be classified in two great families:
\begin{enumerate}
\item {\it The shear models}: these models account for the theory proposed by \citet{Zahn1992}. Examples of these models are the recent
rotating Geneva grids of models \citep{Ekstrom2012, Georgy2013}, as well as the models obtains with the code STAREVOL \citep[for instance the recent grids of models by][]{Charb2010}.
\item {\it The radiative-dynamo models}: these models account for the theory by \citet{Spruit02}.
Examples of such models are the models by \citet{Brott2011, Szecsi2015}.
\end{enumerate}
Shear and radiative-dynamo models predict very different outputs. Shear models predict for instance that stars on the Main-Sequence present some contrast between the angular velocity of the core
and the envelope, the core rotating with an angular velocity a few times higher than the envelope. The  radiative-dynamo models predict a near solid body rotation during the Main-Sequence phase.
For a star with a given initial mass, metallicity, starting its evolution with a given initial velocity, the surface enrichment obtained at a fixed evolutionary stage (for instance at the end of the Main-sequence)
is stronger in radiative dynamo models than in shear models provided no magnetic braking is accounted for at the surface \citep{MMIII2005}. 

The above two families of rotating models can be computed with or without magnetic braking.
Any braking effect, either due to disk locking during the pre-MS phases \citep{Eggen2012}, or due to tides in a close binary systems \citep{Song2013, Song2015}, or due to (wind-)magnetic braking \citep{Meynet2011} have
very different consequences in these two families of models. In case of shear models, any braking boosts the chemical mixing inside the stars predicting stronger surface enrichments than without any braking. In contrary, in radiative-dynamo models, braking 
slows down the mixing process and decreases the surface enrichments \citep{Meynet2011}. Thus, we see that when comparing models to observations, it is important to be aware of these differences between these various families of models. In the rest of the paper,
we discuss results obtained with shear models without magnetic braking.

\begin{figure}
\includegraphics[width=\linewidth,height=85mm]{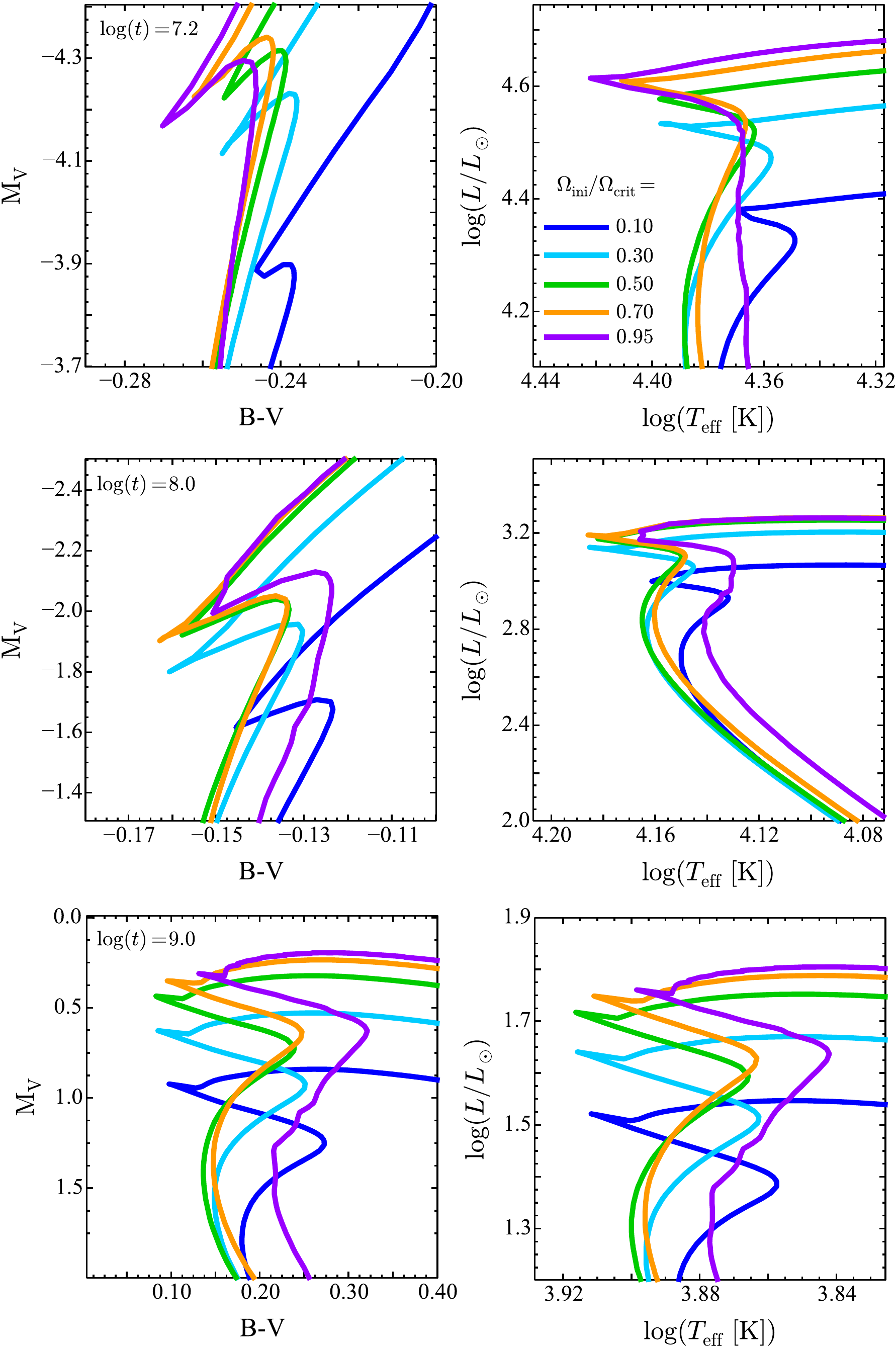}
\caption{Isochrones for various ages computed from stellar models with a Dirac initial distribution of velocity. The initial velocities are indicated on the upper right panel. The mass at the turn off for the logarithm of the  ages of $7.2$, $8.0$ and $9.0$ are around $12$, $5$ and $2 M_\odot$, respectively. This figure was taken from \citet{Georgy2013}.
}
\label{iso}
\end{figure}

\begin{figure}
\includegraphics[width=\linewidth,height=75mm, angle=0]{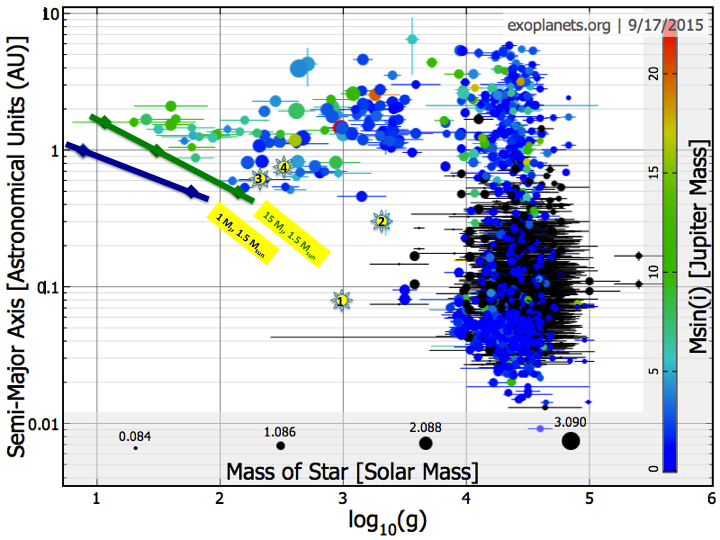}
\caption{Semi-major axis of planet orbits versus the log g of the host star using the database of the exoplanets.org \citep{Han2014}. The colors allow to have an indication on the mass of the planet, while the size of the symbols
are related to the mass of the star. The heavy lines in the upper left corner show the minimum semi-major axis for survival of the planet as computed in the present work. These lines are labeled by the masses of the planet and of the star considered.
The yellow points indicate the positions of the system on the verge of being engulfed: 1= Kepler-91b  \citep{Lillo2014};  2=Kepler-432b \citep{Ortiz2015}; 3=BD+15 2940 \citep{Nowak2013}; 4= HD  233604 \citep{Nowak2013}. 
Figure taken from
Privitera et al. (in preparation).
}
\label{comp}
\end{figure}

\section{Rotation and age determination}

Since rotation affects the evolutionary tracks and the stellar lifetimes, it will change the isochrones. This is illustrated in Fig.~\ref{iso}.
The colored lines connect the positions in the HR diagram of various initial mass stars having all the same age (indicated in the left panels) and the same initial
rotation on the ZAMS. If we consider a real stellar cluster, we shall have stars distributed along these various isochrones with a frequency
depending on the initial distribution of the rotation\footnote{Note that effects of inclinations can be quite large for fast rotating stars. See Figs. 4 and 6 in \citet{Georgy2013}.
}.  We see that the dispersion in visual magnitude (M$_v$) at the turn-off
due to evolutionary effects of rotation amounts to 0.4 mag. respectively for an age equal to 7.2 (in years and in logarithm). The dispersion
increases when larger ages are considered. At an age equal to 9.0, the scatter in M$_v$ amounts to about 0.6. 
This scatter in visual magnitude can be translated into an age scatter when the magnitude at the turn-off is used as an aged indicator.
The age scatter due to the above effect ranges between 25 and 45\% \citep[see Table 2 in][]{Age2009}.

The changes of the isochrones due to rotation produces some widening of the upper main-sequence of stellar clusters mimicking some dispersion in ages when interpreted
from non-rotating stellar models \citep[see the discussion in the recent papers by][]{Bast2015}.
It would be interesting to determine the masses at the turn off of those clusters, since depending whether we have a dispersion of ages, or the same age but a dispersion of 
initial rotation, different mass distributions will be obtained.

\section{Rotation and planets}

Planets may have an impact on the rotation of stars. First,  for allowing planets to form, the proto-planetary disk must last for long enough durations.
Longer disk lifetimes imposes longer disk locking periods and thus stronger braking down of the star in the pre Main-Sequence phase. Thus we could expect that stars with planets will begin their core H-burning phase with smaller initial rotations
and a lower lithium surface abundance \citep{Bouvier2008, Eggen2012}\footnote{The conclusion about lithium depends whether
the stars have internal differential rotation or not.}. Second, due to tidal forces, the semi-major axis of a planet orbit may change, more precisely the tidal forces may produce exchange of angular momentum between the orbit of the planet and the star.  This transfer may be in both ways, either decelerating the planet and making its orbit smaller, or accelerating the planet and making its orbit larger. The first case is obtained when the orbit of the planet is inside the corotation radius, the second when it is outside.
The corotation radius is the radius of the planet orbit such that the period of revolution of the planet is equal to the period of rotation of the star. Third, as a result of the orbital decay described above, planets may be engulfed into the star.

There are some indirect observational evidences that these processes indeed occur in nature. One of them is shown in Fig.~\ref{comp} which plots the semi-major axis of detected planets around stars with initial masses between
about 1 and 3 M$_\odot$ as a function of the surface gravity of the host stars.
Around Main-Sequence stars, {\it i.e.} stars with gravities superior to about 3.8-3.9, the semi-major axis goes from 0.01 AU up to around 7 AU.
Around red-giant stars with gravities below about 3, only planets with semi-major axis between
0.4-0.5 AU up to 5 AU are observed. Of course some observational bias are present, but this is striking that no planets with a semi-major axis inferior to about 0.5 AU are observed around low surface gravity red-giant stars.
Many authors have suggested that this is due to the fact that the planets with smaller semi-major axis have been engulfed by their host stars \citep[see for instance][]{villaver09}. 

This engulfment has the more chance to occur during the red-giant phase where the star
is much larger than on the Main-Sequence and where the energy of the tides is the most efficiently dissipated in its outer convective zone.
Using stellar models, with different initial rotations, we computed the evolution of the orbit for planets of different masses (between 1 and 15 M$_J$),
orbiting stars with initial masses between 1.5 and 2.5 M$_\odot$ and at initial distances ranging from 0.5 up to 5 AU (Privitera et al. in preparation).
The changes of the orbits result not only from the tidal forces between the star and the planet but also from other (less important effects)  like the mass losses of the star, the changes of the mass of the planet by accretion and/or evaporation, the friction and the gravitational drag resulting from the planet moving in a non-empty medium. All these effects have been accounted for.

One result of these computations is that the range of initial radii for which engulfment occurs during the red-giant branch becomes smaller when the initial rotation of the star increases. 
For instance, all 10 M$_J$ planets with an initial orbit radius below 2.6 AU
orbiting 2 M$_\odot$ with an initial angular velocity equal to 10\% the critical velocity will be engulfed before the star reaches its red giant tip. 
When the star has an initial rotation equal to 50\% the critical velocity, only planets with an orbit radius below
2.0 AU will be engulfed. The physical reason for this difference is the following one: first, the maximum radius for engulfment is fixed by the maximum radius of the host star which is reached at the tip of the red-giant branch (this is the last chance for the planet to be engulfed during the first red-giant branch ascent). Second, this maximum radius depends a lot on the fact that the star has or has not a degenerate core at He-ignition. The maximum radius is much larger for stars having a degenerate core than for those
having a non-degenerate one.  This comes from the fact that  the temperature for helium ignition is reached for larger massive cores in degenerate conditions than for non-degenerate one \citep[see e.g. Fig. 14 in][]{MM89}. Since the luminosity increases with the mass of the core, the luminosity and also the radius at the red giant tip will be larger when a degenerate core is present than when a non-degenerate core is present.
Third, our 2 M$_\odot$ happens to be in the transition mass range between the low mass star range where He ignites in degenerate conditions and the intermediate mass range where He ignites in non-degenerate conditions. 
By increasing the convective core during the MS phase, rotation allows the He-core to be less sensitive to degeneracy effects and make possible an He-ignition in less degenerate conditions, thus at a lower luminosity and a smaller radius.
In the domain of mass where there is the transition between He-flash and regular He-ignition, we expect therefore  some scatter in 
the minimum semi-major axis for survival of the planets introduced by
the distribution of initial velocities\footnote{Changing the prescription for the mass loss rate during the red giant branch has an impact on the survival limit too. According to \citet{Villa2014} this impact is however not very large.}.

Are the limits computed compatible with the observed distributions of semi-major axis of planets around red-giant stars? It seems that this is the case. 
In Fig.~\ref{comp}, we have superposed the limits obtained by our computations (see the heavy segments labeled by the planet and stellar masses).
We see that the limits obtained by the computation are relatively near the observed limits at low gravity. This supports the idea
that indeed the planets with smaller initial semi-major axis have been engulfed by the stars during the red giant branch ascent.
Note that during the orbit decay, planets will have semi-major axis below that limit. However there is little chance to see a planet in that stage because, the orbit decay can be seen as a runaway process once it is switched on. Indeed the timescale for the
process varies as $(a/R_*)^8$, where $a$ is the semi-major axis (which decreases during the orbit decay) and $R_*$ the stellar radius, that in general increases along the red giant branch (there is an exception however at the bump).

The process of orbit decay has for consequence to accelerate the convective envelope of the star. We have obtained that this tidal interaction that occurs before the engulfment can already rise the stellar surface velocity
up to larger values than the maximum value that can be expected from internal processes (see Privitera et al. in preparation). Thus tides alone, even in case the planets would be evaporated just before being engulfed, can accelerate the surface rotation of red giants.
We have also checked that these high surface velocities would be observable over relatively long periods, since the internal processes for the redistribution of the angular momentum inside the evolving star as well as the impact
of the changes of radius of the star during its evolution, would not remove rapidly the surface velocity increases due to the orbit decay (Privitera et al. in preparation). Thus this process could account for the existence of the few percents
of fast rotating red giants \citep[see for instance][]{Carl2012}.


\section{Rotation and the early chemical evolution of the Milky Way}

The impacts of rotation are particularly spectacular at low or very low metallicity. This comes from mainly two reasons: first there are some
indications that the number of fast rotators increases when the metallicity decreases \citep{MaederGrebelM, Martayan2007}; second, rotational mixing is more efficient at low metallicity, mainly due to the fact that stars
are more compact \citep{MMVII}. 

At very low metallicity, the mixing induced by rotation allows some exchanges of matter between H-burning and He-burning zones opening the path
to a very rich and interesting nucleosynthesis  \citep[see Table 2 in][]{CEMP2015}. In particular, rotation makes possible primary\footnote{An element is said primary if the quantity that is newly produced by a star does
not depend on the initial metallicity of the star. That means that a Pop III star would produce a primary element in quantities comparable to the quantities produced by a solar-type star.} productions of $^{13}$C, $^{14}$N, $^{22}$Ne \citep{MM2002Nitrogen, MMVIII, Meynet2006, Hirschi2007, Meynet2010}. It also boosts
the production of s-process elements in massive stars at least down to some metallicity limit \citep{Pign2008, Frisch2012}.

The primary production of nitrogen by rotating models can reproduce  the nitrogen to oxygen plateau shown by halo very metal poor stars \citep{Chiappini2006}.
There are at least four other signatures indicating that spinstars, {\it i.e.} stars having their evolution deeply affected by their axial rotation, played a key role in
the early chemical evolution of the Milky Way and more generally in very low metallicity environments \citep[see the review by][and references therein]{Chiappini2013}. 

Among the many interesting topics in this area of research is the case of the Carbon-Enhanced-Metal-Poor (CEMP) stars \citep[see a recent discussion in][]{Norris4} and more particularly of the CEMP-no stars.
The CEMP-no stars are a subclass of the  more general family of the  CEMP stars. They have very low iron contents and present strong overabundances of carbon (as well as of nitrogen and oxygen in general), exactly like the other CEMP stars,  but show no or only very small excesses in s- and r-
elements, hence the {\it no} subscript. The most iron poor stars known so far are CEMP-no stars.
\citet{Meynet2006,Meynet2010, CEMP2015} have argued that these stars might be produced from material ejected by spinstars. The great variety of observed abundances could be due to successive exchanges of material between the Helium burning core and the Hydrogen burning shell in the spinstar \citep{CEMP2015}. 
Interestingly, both the normal halo stars (those that show no strong carbon enhancements) and the CEMP-no stars, might be produced by the same physics operating at different (although low) metallicities, namely the physics of rotational mixing. In both case this mixing is required
by the observed abundances. What might be different is that, while normal halo stars are formed from a relatively well mixed reservoir, enriched by many different types of massive stars, the CEMP-no stars are formed from local pockets of material
enriched by only one or maybe two events. 

One can wonder why the most iron-poor stars are carbon-rich? 
Does this mean that the whole interstellar medium (ISM) was carbon-rich? Probably not, the ISM in the early phase of the galactic halo formation was probably made of different patches of material with different chemical composition depending on the sources that
enrich it. The pieces of ISM that were lacking of heavy elements (iron, carbon..) could not fragment and produce low-mass long-lived stars until they are enriched by other stars, while the piece of ISM enriched at a sufficient high level in carbon could directly fragment and form low-mass long-lived stars that we can still observe today \citep{Gil2013}.
\bibliographystyle{an}
\bibliography{meynet-bib}
%


%
%

\end{document}